\documentclass[preprint]{aastex}

\newcommand{\msun}{M$_{\sun}$}

\begin{document}

\title{Non-thermal Emission in Sagittarius B?}
\author{Cornelia C. Lang}
\affil{Department of Physics \& Astronomy, University of Iowa, Iowa City, IA 52245}
\email{cornelia-lang@uiowa.edu}
\and
\author{Patrick Palmer}
\affil{Department of Astronomy \& Astrophysics, University of Chicago, Chicago, IL 60637}
\and
\author{W. M. Goss}
\affil{National Radio Astronomy Observatory, Socorro, NM 87801}

\begin{abstract}

We summarize three recent publications which suggest that the Galactic
center region Sagittarius B (Sgr B) may contain non-thermal radio components 
(Crocker et al. 2007, Hollis et al. 2007 and Yusef-Zadeh et al. 2007a). Based on new VLA matched-resolution
continuum data at 327 MHz and 1.4 GHz, we find no evidence for large
scale non-thermal radio emission at these frequencies; the spectral
behavior is likely determined by the complex summation of multiple HII
region components with a wide range of emission measures and hence
radio turn-over frequencies.  Also, we discuss a possible additional
interpretation of the radio continuum spectrum of individual component
Sgr B2-F carried out by Yusef-Zadeh et al; confusion from nearby HII
components with widely different turn-over frequencies may contribute
to the the change in slope of the radio continuum in this direction at
low frequencies. Finally, we discuss the uncertainties in the determination 
of the spectral index of the GBT continuum data of Sgr B carried out by Hollis et
al. We find that the apparent spectral index determined by their
procedure is also likely due to a summation over the many diverse
thermal components in this direction.

\end{abstract}

\section{Introduction}

The radio spectrum of the Galactic Center (GC) has been a longstanding and
complicated problem for astronomers.  The GC radio source was slowly recognized
to be a complex of individual sources embedded in diffuse Galactic background emission. 
By 1965, when Bernard Burke's review article, entitled "Radio Radiation
from the Galactic Nuclear Environment", appeared in the {\it Annual Review of
Astronomy and Astrophysics}, images of the GC region had been made
at a range of frequencies as high as 14 GHz. One of the images in
this review article was from Cooper \& Price (1964), made with the Parkes,
Australia, 210-foot telescope at 3 GHz. This 
image had a resolution of 6.7\arcmin, making it possible to resolve the 
Sagittarius B (Sgr B) complex from the other major 
GC sources. Cooper \& Price (1964) were the first to suggest that the Sgr B source is an optically-thick 
thermal source based on a spectral index measurement between 3 and 8 GHz (Drake 1959a,b) 
that had large uncertainties.
During the last 40 years, the Sgr B region has been the subject of numerous detailed studies
carried out with the increasing resolution and sensitivity of new instrumentation.

\subsection{The SgrB Region: Sgr B2 and Sgr B1}

The Sgr B region is now understood to be one of the most complex star-forming regions in the
Galaxy.  Located at $\sim$105 pc in projection from Sgr A$^*$, the Sgr B giant
molecular cloud has a total mass of 8-20 x 10$^6$ \msun~(Tsuboi et
al. 1999). Therefore, many interstellar molecules have been detected
for the first time in the Sgr B cloud. Most studies of Sgr B have
been carried out in the far-infrared and radio regimes because this region
of the Galaxy is highly obscured (by up to 50 magnitudes) at visible
wavelengths. 

The Sgr B region is comprised of two distinct complexes:  Sgr B2 (G0.7-0.0; to the
North) and Sgr B1 (G0.5-0.0; to the South). Both sources are strong radio continuum
sources and both have been well-studied on fine scales ($\sim$1\arcsec~or
better) with interferometric observations.  In higher frequency
radio observations ($\nu$ $>$ 1.4 GHz), Sgr B2 dominates due to the numerous
ultra-compact, optically-thick HII regions (Benson \& Johnston 1984; Gaume
\& Claussen 1990). 

Sgr B2 is well-known for its incredibly high densities ($>$ 10$^5$
cm$^{-3}$) in both ionized and molecular gas (Huttemeister et al. 1993;
DePree, Goss \& Gaume 1998).  The Sgr B2 complex has three bright cores -
Sgr B2 ``Main'', ``North'' and ``South''. Sgr B2 shows radio emission over a
factor of 6000 in size scale, from $<$1\arcsec~(ultra-compact HII regions) 
to the 10\arcmin~extent of diffuse emission along the complex. 
The highest resolution radio study of Sgr B2 was made at 43 GHz with the
Very Large Array (VLA)\footnotemark\footnotetext{The
 National Radio Astronomy Observatory is a facility of the National Science Foundation operated under cooperative agreement by Associated Universities, Inc.} radio telescope with an angular resolution of 
$\sim$64 milli-arcseconds (600 AU at the GC distance of 8 kpc) in SgrB2 Main. 
This study revealed over 20 ultra-compact HII regions and stellar wind
sources in Sgr B2 Main alone (De Pree et al. 1998). In addition, the range of observed emission
measures in the ionized gas are from 1$\times$10$^4$ to 1$\times$10$^9$ pc
cm$^{-6}$.  Radio recombination line studies at numerous frequencies have 
helped cement the thermal nature of many of the individual radio components
in the Sgr B complex (e.g., Mehringer et al. 1993). 
The radio emission arising from the Sgr B1 complex
indicates that this region is a more highly evolved
star-forming region with numerous filamentary and shell-like ionized
structures (Mehringer et al. 1992). Lying between Sgr B1 and Sgr B2 is a complex of
radio sources known as G0.6-0.0. Gas velocities and morphology suggest that
G0.6-0.0 physically connects Sgr B1 and Sgr B2 (Mehringer et al. 1992). 

\subsection{Recent Papers: Non-thermal Emission in Sgr B?}

Three recent papers on Sgr B complex have addressed the nature of the radio
emission arising from this region, and have argued that there is a
non-thermal component of the radio emission (Crocker et al. 2007; Hollis et
al. 2007; and Yusef-Zadeh et al. 2007a). Because Sgr B is one of the 
most active regions of star formation in the Galaxy, containing more than 60 HII regions,
it would not be surprising to find an embedded supernova remnant or two.  However, because 
the preponderance of evidence for the last four decades has indicated that the 
Sgr B complex is a thermal radio source, claims of non-thermal emission deserve close
scrutiny.  

One of the main motivations for looking for non-thermal emission has come
from high-energy investigations of this region. Using HESS, Aharonian et al. 
(1996) have made a detection of diffuse $\gamma$-ray flux between 0.2-20 TeV (1 TeV = 10$^{12}$
eV) from the Galactic center (GC) region, distributed along the Galactic
plane. In addition, a separate HESS measurement of the
$\gamma$-ray flux and spectrum in a 0.5\arcdeg$\times$0.5\arcdeg~region
was centered on the SgrB molecular cloud.
After removal of several bright point-like sources, the diffuse flux is believed to be 
correlated with the density of molecular gas
(H$_2$) in the GC region, which has been imaged in the CS (J=1-0)
line transition (Tsuboi et al. 1999). One possible origin of the high-energy radiation is the 
collision between cosmic ray protons and the ambient, dense molecular gas in the GC.  
Crocker et al. (2007) predict the radio synchrotron spectrum from their broadband (radio to $\gamma$-ray) 
emission models and compare their results to the measured radio spectrum of
this region.  Crocker et al. (2007) base their radio spectrum of this region primarily on
low-frequency data at 327 and 843 MHz (see below for a description). They
determine that the radio spectrum of Sgr B is non-thermal in nature, with a measured excess
of non-thermal emission.

In addition, Yusef-Zadeh et al. (2007a) consider the global heating of
molecular clouds by cosmic rays, based on their earlier observations of correlations
between molecular gas and 6.4 keV K$\alpha$ line emission (Yusef-Zadeh et al. 2007b). 
In order to illustrate that SgrB2 has a non-thermal
component, Yusef-Zadeh et al. (2007a) use low frequency observations
between 255 MHz and 1.4 GHz to show that a part of the SgrB2 complex (a
well-studied part, SgrB2 ``F'', a cluster of ultra-compact HII regions) has
a component of non-thermal emission with a flux density of $\sim$82 mJy.
They suggest the presence of enhanced heating by
cosmic ray particles (which increases the ionization fraction) in the
molecular gas. 

Finally, Hollis et al. (2007) 
present a study of the continuum temperature of Sgr B2 based on GBT spectral line observations. They
argue that there is a non-thermal component in Sgr B2 on size scales of
$\sim$143\arcsec~at 1.4 GHz with an optically-thin spectral index of $\alpha$=$-$0.7.

In this paper, a discussion of the measurements used to derive the spectral index 
of radio emission in the above papers follows. In addition, a high-resolution
1.4 GHz image of the GC region is presented and used for comparison to the three
recent results on the non-thermal emission in Sgr B. 

\section{Radio Observations of the Sgr B Region at Low Frequencies}

\subsection{1.4 GHz VLA Image}

Figure 1 shows a 1.4 GHz VLA image of the Sgr B region of the GC. This image was made
using the DnC and CnB array configurations as part of an HI absorption
study toward the central 250 pc of the Galaxy (e.g., Lang et al. (2003) and
Lang et al.  {\it in prep.}). In order to cover this large region, 5 fields were mosaicked
together using the miriad task {\it mosmem}. The integration time was 8
hours for the set of five fields in each configuration (i.e., approximately
2.5 hours on each field) which resulted in an rms of 10 mJy beam$^{-1}$. 
The zero level in this image is within 1$\sigma$ (10 mJy) of zero, and the image has 
also been corrected for the variable system temperature with frequency in the presence 
of HI emission. This mosaic is one of the highest-resolution images of the more extended
structures in the complex GC region, with excellent sensitivity to both point-like and
diffuse features.  In this paper, this image is used to determine the flux density of 
the Sgr B complex and to help constrain the radio spectrum of Sgr B. 

\begin{figure}
\plotone{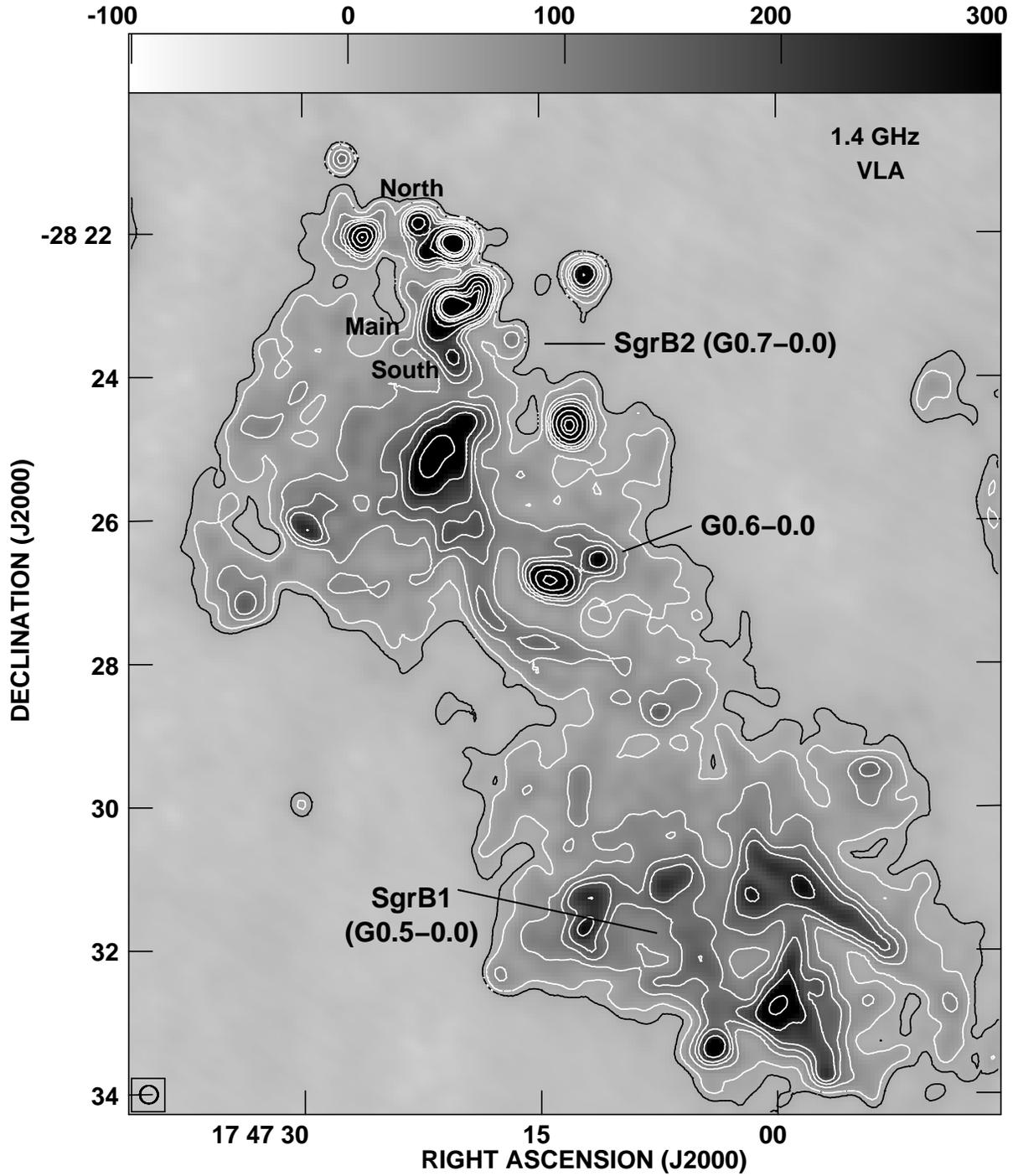}
\caption{VLA 1.4 GHz continuum image of the SgrB1 and B2 region from Lang et al. (2003) 
and Lang et al. {\it in prep.}. The image has a spatial resolution of 15\arcsec~$\times$
15\arcsec~and an rms level of 10 mJy beam$^{-1}$. Contours are at levels of 3, 5,10, 15, 20, 30, 
50, 90, 100, and 125 times the rms level.}
\end{figure}

\subsection{Flux Density of the Sgr B Complex at Low Frequencies}

Based on data at 327 and 843 MHz, Crocker et al. (2007) 
determine that the radio emission arising from the Sgr B region is non-thermal. 
At 327 MHz, Crocker et al. (2007) use data from Brogan et al. (2003) smoothed from the
original resolution of 120\arcsec~$\times$ 60\arcsec~to 252\arcsec. They measure the integrated flux 
density in two regions: (1) the field of view of the HESS detection, which is a large
region of 0.5\arcdeg~$\times$0.5\arcdeg~size centered on Sgr B and (2) a smaller region 
containing primarily the bright radio emission associated with Sgr B1 and B2 radio sources. 
For the large region, they measure the flux density to be 45$\pm$10 Jy, and for the smaller region they 
measure 25 Jy.  At 843 MHz (using an image from SUMSS; Bock et al. 1999), 
a two sigma detection of 20$\pm$10 Jy for the 0.5\arcdeg~$\times$0.5\arcdeg~region is 
obtained by Crocker et al. (2007). They also attempt to measure the flux density at 74 MHz in an
image from Brogan et al. (2003), but they can not obtain a result for
the flux density of this region. It is likely that the spectrum derived by Crocker et al. 
(2007) is not reliable due primarily to the poorly-determined flux density at 327 MHz, 
coupled with the poor signal-to-noise of the 843 MHz data. At 327 MHz, there are large 
distortions in the image due to missing flux and confusion from SgrA West and East. 
It is not possible to obtain an accurate flux density over a region as large 
as $\sim$0.5 \arcdeg~$\times$ 0.5\arcdeg.

We suggest that a more reliable radio spectrum can be determined by using 
VLA data at two frequencies (1.4 GHz and 327 MHz) with matched {\it (u,v)} coverage.
Data from Brogan et al. (2003) were used at 327 MHz and at 1.4 GHz from Lang et al. 
(2003; also presented above). The flux densities were obtained over a region which
includes the radio emission associated with Sgr B1 and B2 (up to an angular size of 
$\sim$0.3\arcdeg), with careful consideration of the problems of the displacement of the zero level. At
both frequencies, good signal-to-noise determinations are possible. We have used the
AIPS task {\it IRING} to determine the flux in concentric rings. The total cumulative flux density 
can then be determined as a function of radius. The total flux density at 1.4 GHz  
over a region that includes both Sgr B1 and B2 is 83$\pm$7 Jy with a radius of 500\arcsec~(8.3\arcmin). 
For Sgr B1 and B2 individually, we determine separate flux densities at 1.4 GHz 
of 35$\pm$5 Jy and 49$\pm$6 Jy. At 327 MHz, we determine a flux 
density of 37$\pm$7 Jy for Sgr B over a region of the same region used at 1.4 GHz. 

In Table 1 we summarize the flux density determinations from this paper and compare
these with Crocker et al. (2007) and previous VLA measurements. 
The agreement with Mehringer et al. (1993) is encouraging. In addition, the 327 MHz measurement for Sgr B2 of
Yusef-Zadeh et al. (2007b) agrees well with our determination (see Table 1). 
Our results, which have high signal-to-noise and matched {\it (u,v)} coverage, provide
good evidence that the flux density of the Sgr B complex {\it increases} from 327 MHz
to 1.4 GHz, in contrast to the conclusion of Crocker et al. (2007). The current data are in 
good agreement with previous independent VLA studies (see Table 1 and Mehringer et al. 1992, 1993).  

A likely explanation of the radio spectrum of Sgr B we derive (i.e., thermal) is the large 
halo of size 10\arcmin$\times$5\arcmin~first described by Mehringer et al. (1993). 
This halo represents a ``transition'' region from HII gas optically thick at 327 MHz
becoming optically thin at the higher frequencies. In order to provide a crude estimate of
the physical parameters of the halo, we estimate that $\sim$50\% of the emission at 1.4 
GHz arises from the halo. This estimate is derived by summing all the compact components
in Figure 1 ($\sim$40 Jy) and subtracting this from the total flux density of $\sim$83 Jy 
(see Table 1) to obtain a flux density of $\sim$40 Jy. For a source of radius $\sim$15 pc,
the emission measure is 1$\times$10$^5$ pc$^{-6}$, with an rms electron density of 60 cm$^{-3}$.  
The implied opacity in the continuum at 327 MHz is then $\tau$$\sim$0.3 for a source
of this size. Thus, the ratio of flux densities from 1.4 GHz to 327 MHz for the entire
Sgr B complex is 2.2 and not the ratio for a source that is completely optically thick
at both frequencies (which would be 21). This small ratio of flux densities ($\sim$2) 
is likely due to a very gradual turnover in the spectrum at these frequencies because of 
the wide range of emission measures across the Sgr B complex. 

Finally, we note that Yusef-Zadeh et al. (2007a) determine that a small region
within the Sgr B2 complex is non-thermal, Sgr B2-F. The cluster of
ultra-compact HII regions known as Sgr B2-F has been studied at very high angular
resolution and measures only several arcseconds across (DePree et. al. 1996).
Yusef-Zadeh et al. (2007a) suggest that the flux density at 255 MHz arising from this region (Sgr
B2-F) is enhanced due to a non-thermal contribution of this source. However, an
alternative interpretation is that the excess flux density that Yusef-Zadeh et al. (2007a) measure at 255
MHz may be due to confusion from the numerous sources within the GMRT beam (22\arcsec).
There are likely to be more than 25 compact sources within the 22\arcsec~beam, based on
Figures 2 and 4 in Gaume et al. (1995). For example, source Sgr B2-I lies only
$\sim$4\arcsec to the E of Sgr B2-F and has a flux density in the range of 100-200 mJy at 1.375 MHz (Mehringer et al.
1995). Some of these sources may well have lower density structures which will contribute to a
complex distribution of opacities and emission measures within the GMRT beam, resulting in 
a slight deviation of the spectrum from the optically thick portion of the low frequency spectrum. 

\begin{deluxetable}{cccccccc}
\tabletypesize{\small}
\tablecolumns{8} 
\tablewidth{0pc} 
\tablecaption{Radio Flux Density of the Sgr B Complex}
\tablehead{
\colhead{Frequency}& 
\colhead{VLA}&
\colhead{Spatial}& 
\colhead{Sgr B} &
\colhead{Sgr B1}       &
\colhead{Sgr B2}      & 
\colhead{Reference} \\
\colhead{(GHz)} &
\colhead{Array(s)} &
\colhead{Resolution}&
\colhead{(Jy)} &
\colhead{(Jy)} &
\colhead{(Jy)} & 
\colhead{}}
\startdata
1.4 &DnC,CnB &15\arcsec$\times$15\arcsec &83$\pm$7 &35$\pm$5 & 49$\pm$6\tablenotemark{*}& This paper\\
1.4 &D,C&26\arcsec$\times$15\arcsec &75 & 26 & 39\tablenotemark{*}& Mehringer et al. (1993)\\
\hline
0.843 & $--$ & & 20$\pm$10\tablenotemark{**} &$--$ &$--$ & Crocker et al. (2007)\tablenotemark{\ddag}\\
\hline
0.327 & C,D & 60\arcsec$\times$120\arcsec& 37$\pm$7&$--$& 17$\pm$4& This paper\tablenotemark{\dag}\\
0.327 & B,C,D & 41.6\arcsec$\times$22.7\arcsec&$--$&$--$ &17& Yusef-Zadeh et al. (2007b)\\
0.327 & C,D & 252\arcsec$\times$252\arcsec&45$\pm$10\tablenotemark{**}&$--$ &$--$ & Crocker et al. (2007)\tablenotemark{\dag}\\
\hline

\enddata
\tablenotetext{*}{Sgr B2 region includes emission from the G0.6-0.0 complex.}
\tablenotetext{**}{Flux density is for the HESS large region of size 0.5\arcdeg$\times$0.5\arcdeg.}
\tablenotetext{\dag}{Data from Brogan et al. (2003).}
\tablenotetext{\ddag}{Data from Bock et al. (1999).}
\end{deluxetable}

\subsection{Continuum Temperature of Sgr B2}

Hollis et al. (2007) noted that the Green Bank Telescope (GBT) 
continuum antenna temperature toward Sgr B2 is linear on a log-log plot.  From this
straightforward result, they carried out an analysis suggesting that 
the bulk of the emission from this direction is non-thermal and
can be associated with a source of size 143" at 1 GHz, with a source
size decreasing with frequency as $\nu$$^{-0.52}$.  We examine their methods
and results, and suggest some problems in their analysis. We address three
points: the effect of source structure on the measured continuum temperature, the
estimates of source size from archival VLA observations, and the question of
the 43 GHz flux density from this region. 

At their lowest frequency (1.35 GHz), the FWHM of the GBT beam is
9.1\arcmin, while at their highest frequency (47.68 GHz) the FWHM is
16\arcsec.  Within a FWHM of 9.1\arcmin, more than 100 components in the Sgr B2 
complex are detected in high resolution images. The majority of these 
components are known to be thermal both because of their radio continuum spectra 
and because of the presence of recombination lines (Gaume \& Claussen 1990, 
Mehringer et al. 1992, 1993).  Emission measures of these components
have a very wide range of values.  Because the GBT beam scales as $\nu^{-1}$, these components will contribute 
differently to the measured flux density due to their relative positions 
and to the changes in opacities because of the scaling of $\tau$ as $\sim \nu^{-2.1}$. 
Therefore, the apparent non-thermal spectral index ($\alpha$=$-$0.7) of the continuum temperature
measured with the GBT by Hollis et al. (2007) may be due to integrating over the large number of
components in Sgr B2. The opacities change with frequency and the 
contribution of each compact source varies systematically as the GBT beam size changes with
frequency. 

If the flux density of a point source were independent of frequency, then the procedure 
of Hollis et al. (2007) would yield a spectral index of zero.  For an extended source (size $>>$ than
the GBT beam), the measured flux density would scale as the beam area,
i.e. as $\nu^{-2}$.  The measured flux density is proportional to the
quantities plotted in Figures 1a, 1b, and 2a of Hollis et al. (2007) (with minor
corrections for efficiencies).  In their equation (3), a
structure correction (1+ $\theta_{s}^2$/$\theta_b^2$) is introduced to
compensate for the resolution of the source.  This correction would remove
the $\nu^{-2}$ dependence that results from the beam area when $\theta_{s}^2$/$\theta_b^2$ $>>$ 1.

It is useful to demonstrate the effect of integration over a complex distribution.
The 1.4 GHz image shown in Figure 1 is multiplied by the GBT beam pattern (approximated by a
Gaussian with the same FWHM and positioned at the Hollis et al. (2007) pointing
center) and summed to obtain the total flux density ``observed''. This method can be 
carried out at a number of frequencies.  This procedure then shows that an apparent spectral
index can be caused by structure alone. If the flux density is plotted on a log-log plot
against frequency, the result is linear with a slope of $-$1.07 (+/- 0.05).  
This result may be compared with Hollis et al. Fig 1b of the continuum antenna temperature
which has a slope of $-$1.06.  In view of the complicated structure of the
source shown in the Sgr B2 complex in Figure 1, the approximation used by Hollis et al. to make
the structure correction is clearly questionable.  However, if we ignore this misgiving and 
the Hollis et al. (2007) structure correction were made, the slope would become
$-$0.82$\pm$0.08.  This value may be compared with Figure 2b of Hollis et al. (2007) 
in which the slope is given as $-$0.7. While this is not a simulation because we use
data at only one frequency, this demonstration convinces us that the effect of
structure cannot be ignored when trying to interpret the GBT continuum temperature. In our
opinion, the apparent non-thermal spectrum reported by Hollis et al. (2007) is
an effect of source structure alone. 

In addition, estimates of source size obtained by
heavily tapering VLA data are not realistic.  The most compact VLA
configurations were used in all of the archival observations analyzed.  The
shortest spacings (measured in wavelengths) increase with frequency so that
the largest angular scale observable decreases with increasing frequency.
Therefore it is not surprising that the apparent source size decreases with
increasing frequency.  Typically, if VLA data has been tapered so
that the beam size is increased by more than a factor of two to four, the
imaging quality is too poor to be useful. Hollis et al. (2007) acknowledge that this
effect may cause an unknown systematic uncertainty.

Finally, Hollis et al. (2007) indicate that their 43 GHz results are
not consistent with the data of Mehringer \& Menten (1997). 
The GBT pointing center is offset by more than 1/2 the FWHM of
the beam at 44 GHz from the centroid of the Sgr B2 N complex as shown in Figure 1 of 
Mehringer \& Menten (1997). Therefore, the GBT would be expected to measure a lower 
flux density. The four highest frequency points in Figure 2 of Hollis et al. (2007) 
show a systematic decrease in flux density by a factor of $\sim$1.6.  For a
source beyond the half-power point of the beam, the flux density is
expected to decrease rapidly with increasing frequency because of the
simultaneous narrowing of the antenna beam.  Therefore, because of 
the different pointing centers, the 43 GHz data of Mehringer \&
Menten (1997) and that of Hollis et al. (2007) may well be consistent 

In summary, the apparent non-thermal power law slope for the Sgr B2 continuum
temperature observed by the GBT is likely determined by source
structure and provides limited information about the physical processes in
the Sgr B region.

\subsection{Previous Studies of Non-thermal Emission in Sgr B}

The three 2007 papers addressed above are not the first which have suggested a
possible non-thermal component in the Sgr B region.  For example, Reich et al (1987)
compared images of Galactic center sources at 60 $\mu$m and 
2.6 GHz with a resolution of 6\arcmin, and concluded 
that Sgr B2 was possibly non-thermal. Akabane et al. (1988) compared images at 43 and 10.7 GHz 
(convolved to 1.3\arcmin) and suggested possible non-thermal emission 
in the southern part of Sgr B2.  Akabane et al. (1988) also
noted that their result may imply that each of their compact sources may
consist of a complex of HII regions, some optically thick and some
optically thin.  Because of the source structure apparent in Figure 1 and the 
subsequent studies of Sgr B2 with higher angular resolution (DePree et al. 1996, 1998), 
we believe this suggestion is correct.  Haynes et al. (1992) imaged the polarized emission near the
GC with $\sim$2.8\arcmin~resolution.  While one of the arcs of
polarized emission (major axis size of $\sim$0.8\arcdeg) crosses Sgr B2, it would seem to
be a part of the diffuse Galactic emission, not a part of Sgr B2.  

\section{Conclusions}

In this paper, we have summarized three recent papers which point out
possible non-thermal radio emission arising from the Sgr B region in the
GC. We also present a high-resolution and sensitive VLA image of the 
Sgr B region at 1.4 GHz. Using this image and a matched-array 327 MHz 
VLA image, we derive a thermal spectrum for the Sgr B complex and 
suggest that the radio emission is a mixture of optically thin and optically
thick emission over the frequency range discussed here (255 MHz to 1.4 GHz).  
In addition, we show that the the apparent non-thermal power law slope for the Sgr B2 continuum
temperature observed by the GBT is likely determined by source
structure and provides limited information about the physical processes in
the Sgr B region. While the structure Sgr B region is complex and furthermore
confused by the Galactic background, there does not appear to be substantial 
evidence for a non-thermal component in the Sgr B complex. 

The authors would like to thank Roland Crocker, Mike Hollis, and Farhad Yusef-Zadeh 
for their comments on this article.


\begin{references}
Aharonian, F., et al.\ 2006, \nat, 439, 695 \\
Akabane, K., Sofue, Y., Hirabayashi, H., Morimoto, M., Inoue, M. 1988, PASJ, 40, 459\\
Benson, J.~M., \& Johnston, K.~J.\ 1984, \apj, 277, 181 \\
Bock, D.~C.-J., Large, M.~I., \& Sadler, E.~M.\ 1999, \aj, 117, 1578 \\
Brogan, C.~L., Nord, M., Kassim, N., Lazio, J., \& Anantharamaiah, K.\ 2003, 
Astronomische Nachrichten Supplement, 324, 17\\
Burke, B. F. 1965, ARA\&A, 3, 275\\
Cooper, B. \& Price, M. 1964, IAU Symposium No. 20, {\it The Galaxy and the Magellanic Clouds}, 20, 168\\ 
Crocker, R. M., Jones, D., Protheroe, R. J., Ott, J., Ekers, R., Melia, F.,
Stanev, T., Green, A. 2007, ApJ, 666, 934\\
De Pree, C.~G., Gaume, R.~A., Goss, W.~M., \& Claussen, M.~J.\ 1996, ApJ, 464, 788 \\
De Pree, C. G., Goss, W. M. \& Gaume, R. A. 1998, ApJ, 500, 847\\
Drake, F. 1959a, Sky and Telescope, {\it Radio resolution of the galactic nucleus}, June, Vol. 18, 428\\ 
Drake, F. 1959b, Annual Report of the National Radio Astronomy Observatory, July 1, 2\\ 
Gaume, R.~A., \& Claussen, M.~J.\ 1990, ApJ, 351, 538 \\
Gaume, R. A., Claussen, M.~J., de Pree, C.~G., Goss, W.~M., \& Mehringer, D.~M.\ 1995, ApJ, 449, 
663\\ 
Haynes, R. F., Stewart, R. T., Gray, A. D., Reich, W., Reich, P., Mebold, U.
1992, A\&A, 264, 500\\
Hollis, J. M., Jewell, P. R., Remijan, A. J., Lovas, F. J. 2007, ApJ,
660, L125\\
Huttemeister, S., Wilson, T.~L., Henkel, C., \& Mauersberger, R.\ 1993, A\&A, 276, 445 \\
Lang, C.~C., Cyganowski, C., Goss, W.~M., \& Zhao, J.-H.\ 2003, Astronomische Nachrichten Supplement, 324, 1\\
Mehringer, D.~M., \& Menten, K.~M.\ 1997, ApJS, 474, 346 \\
Mehringer, D.~M., De Pree, C.~G., Gaume, R.~A., Goss, W.~M., \& Claussen, M.~J.\ 1995, ApJL, 442, L29\\ 
Mehringer, D.~M., Palmer, P., Goss, W.~M., \& Yusef-Zadeh, F.\ 1993, ApJ, 412, 684\\ 
Mehringer, D.~M., Yusef-Zadeh, F., Palmer, P., \& Goss, W.~M.\ 1992, ApJ, 401, 168\\ 
Reich, W., Sofue, Y., Fuerst, E. 1987, PASJ, 39, 573\\
Tsuboi, M., Handa, T., \& Ukita, N.\ 1999, ApJS, 120, 1 \\
Yusef-Zadeh, F., Wardle, M., Roy, S. 2007a, ApJ, 665, L123\\
Yusef-Zadeh, F., Muno, M., Wardle, M., \& Lis, D.~C.\ 2007b, ApJ, 656, 847 \\
\end{references}
\end{document}